\pdfoutput=1

\documentclass{vgtc}                           

\graphicspath{{figures/}{./}} 

\usepackage{times}                     
\usepackage{mathptmx}                  

\usepackage{booktabs}
\usepackage{multirow}
\usepackage{tabularx}
\usepackage{enumitem}
\usepackage{cite}                      

\newcommand{\Description}[2][]{}

\onlineid{0}

\vgtccategory{Research}

\vgtcinsertpkg

\title{Better Situational Awareness in AR-HRC? A Comparative Study of Augmented Reality and Mobile Interfaces for Human-Robot Collaboration}

\author{Zhehan Qu\thanks{e-mail: zhehan.qu@duke.edu. Zhehan Qu and Christian Fronk contributed equally to this research.}\\ %
        \scriptsize Duke University 
\and Christian Fronk\thanks{e-mail: christian.fronk@duke.edu}\\ %
        \scriptsize Duke University %
\and Jaewoong Jeong\thanks{e-mail: jaewoong.jeong@duke.edu}\\ %
        \scriptsize Duke University %
\and Pavel Manakhov\thanks{e-mail: p.manakhov@lancaster.ac.uk}\\ %
        \scriptsize Lancaster University %
\and Maria Gorlatova\thanks{e-mail: maria.gorlatova@duke.edu}\\ %
        \scriptsize Duke University}

\teaser{
  \centering
  \includegraphics[width=0.98\linewidth]{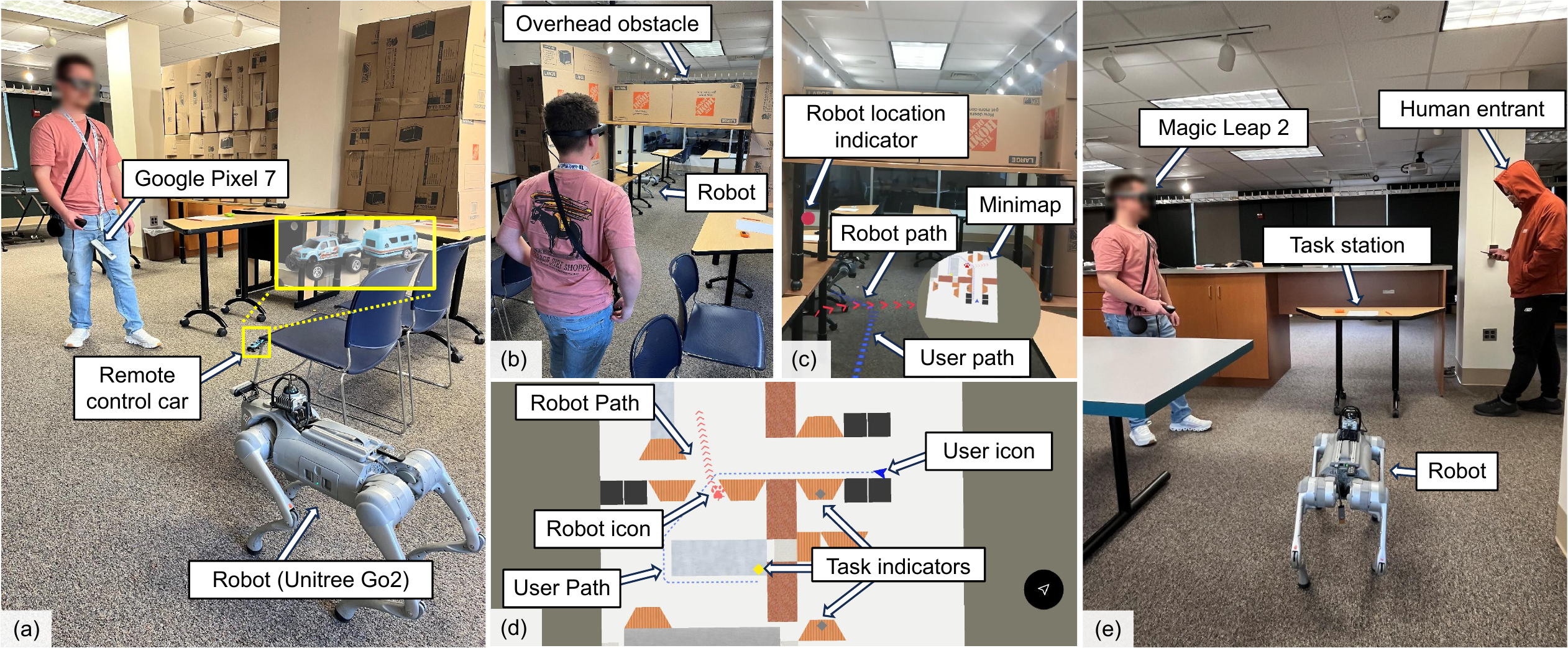}
  \caption{Comparison of situational awareness (SA) in an emulated human-robot collaborative search-and-rescue task, guided by an augmented reality (AR) vs.~a mobile phone interface. SA toward the robot and the environment was measured across three probes: (a) Probe 1: potential collision with the robot and a remote-control car simulating a dynamic environmental hazard; (b) Probe 2: awareness of the robot in the other half of the room and of the overhead obstacle along the user's path; (c) the AR interface and (d) the mobile interface at Probe 2, presenting equivalent robot information and guidance to the next task station; (e) Probe 3: awareness of the robot's heading and next target, and of a human entrant walking by. Note that the robot was intentionally made visible in these images for illustration; these images do not reflect the actual visibility of the robot during the study.}
  \Description{A multi-panel image illustrating the study's three situational awareness (SA) probes in a physical room. Panel (a) shows Probe 1, with a user wearing a headset and holding a Google Pixel 7 mobile phone navigating near a desk while a robot dog and an RC car cross paths nearby. Panel (b) shows Probe 2, with the user approaching a head-level hanging obstacle. Panel (c) displays the first-person AR headset view at Probe 2, featuring a minimap, a conformal blue path leading to the next subtask station and a conformal path composed of red arrows extending from the robot. Panel (d) shows the equivalent Mobile interface at Probe 2 on a smartphone screen, rendering the top-down map. Panel (e) shows Probe 3, where the user encounters another human walking into the task area while the robot heads toward the next subtask station it will scan.}
  \label{fig:teaser}
}

\abstract{
  Augmented reality (AR) facilitates human-robot collaboration (HRC) by enabling in-situ spatial visualizations of the robot and the joint task. However, in safety-critical HRC scenarios such as search-and-rescue, spatial visualizations may also reshape visual attention in ways that create competing situational awareness (SA) demands, potentially introducing new safety concerns. While prior AR-HRC work suggests potential benefits for SA, rigorous evaluations that jointly consider robot and environmental awareness across multiple levels of SA remain limited. We address this through a between-subjects study with 30 participants comparing custom AR and mobile interfaces presenting equivalent information, measuring robot and environmental SA with the Situation Awareness Global Assessment Technique (SAGAT) across all three levels, with concurrent eye tracking to identify the attentional mechanisms underlying any SA differences. Both interfaces achieved high usability; relative to the mobile baseline, AR improved perception-level awareness of the robot but yielded no gains in higher-level robot awareness or in environmental awareness at any level. Gaze analysis explained this: AR freed attention from the map, but that attention was re-invested in the conformal visuals rather than the physical environment. Freeing the eyes from a screen is not the same as directing them to the world, a distinction AR interfaces for safety-critical HRC must design around.

}


\begin{document}
\raggedbottom

\firstsection{Introduction}

\maketitle

Augmented reality (AR) has emerged as a promising interface paradigm for human-robot collaboration (HRC), enabling interaction with robots through spatially grounded, in-situ visualizations. These visualizations can intuitively communicate robot state and intent, as well as workspace constraints and task objectives. Prior work has demonstrated that AR visuals benefit user understanding~\cite{ISMAR_23_Convey_Robot_State_Intent,ICRA_2022_Convey_Arm_Intent,HRI_26_Communicate_Robot_Ability_with_AR_Affordances, IJSR_26_Indicate_Robot_Vision_With_AR}, safety~\cite{HRI_2023_Robot_SafetyZones_AR,ROMAN_2022_Virtual_Barriers}, and performance~\cite{ROMAN_2021_AR_Improve_HRC_Efficiency,ICRA_2021_ARROCH_AR_Robot_Collab_Indoors_TabletBased,IROS_2023_Visualization_Effect_On_Eff,UIST_2022_AR_Program_TTRobots,UIST_2024_AR_Robot_Training}. However, for AR-HRC interfaces to support real-world collaboration, they must do more than convey information alone: they must also preserve users' situational awareness (SA), or their ability to perceive relevant physical elements, comprehend their meaning, and anticipate their future states. SA is critical for safe and effective coordination in HRC because users must simultaneously monitor the robot and interpret the surrounding environment. This challenge is particularly evident in safety-critical domains such as search-and-rescue (SAR), where the environment changes over time and users have only partial information about it. Users need to combine information from different locations, respond to unexpected events, and anticipate changes in robot behavior and the environment.

At first glance, spatially conformal AR seems well-suited to address these SA demands. Prior work suggests that grounding virtual cues in the physical workspace supports more effective attention division than screen-based displays~\cite{Wickensetal2022}, allowing users to monitor the robot without looking away from the environment. However, keeping the environment in view does not guarantee that users will notice relevant events. Visually prominent conformal AR guidance may draw users' attention away from hazards, obstacles, and changes in the environment that they might otherwise notice with a conventional interface. Whether AR improves overall SA in dynamic, safety-critical HRC settings therefore remains unclear.

Prior AR-HRC evaluations that claim SA benefits often substantiate those claims with usability, task performance, or robot-only SA measures~\cite{Telepresence_2024_AR_Enhance_SA_in_HRI,chacko2026making}, with limited direct evaluation of SA using structured frameworks~\cite{endsley1995toward,endsley1988situation} that distinguish among different targets and levels of SA. Moreover, existing evaluations often involve stationary robots, static virtual setups, or tasks with few salient events~\cite{ROMAN_2023_Effect_of_AR_on_SA_in_HRC,chacko2026making}. These conditions may not capture the changes, occlusions, and limited visibility that users encounter in SAR. They also offer limited insight into how AR compares with standard interfaces, such as mobile phones, when both present the same information. As a result, it is difficult to determine whether observed benefits arise from AR's spatial visualizations or simply from access to additional robot and task information. Finally, SA scores alone reveal \textit{whether} awareness differs between interfaces, but not \textit{why}. Examining how users allocate their attention may help explain these differences and inform interface design.


To address these gaps, we present an IRB-approved, between-subjects user study with 30 participants comparing a custom AR interface with an information-equivalent mobile baseline in a SAR-inspired HRC task involving a Unitree Go2 quadruped. The study incorporates dynamic hazards and constrained visibility characteristic of SAR environments and centers on two research questions: \textbf{RQ1}, how AR affects users' SA toward the robot \textbf{(RQ1a)} and the surrounding environment \textbf{(RQ1b)} across the perception, comprehension, and projection levels; and \textbf{RQ2}, how AR reshapes users' allocation of visual attention, and whether these attentional differences explain its effects on SA. We answer RQ1 using SAGAT~\cite{endsley1988situation} probes administered during the task, and RQ2 by analyzing participants' eye movements and linking gaze behavior to SA outcomes through mediation analysis.

Overall, our results show that AR's benefits for HRC are selective, and that gaze behavior helps explain both where they appear and where they do not. We use these findings to derive recommendations for AR interface design. Our contributions are as follows:
\begin{itemize}[itemsep=0pt, topsep=0.3pt, parsep=1pt, leftmargin=0.15in]
    \item We designed a comprehensive SAR-inspired AR-HRC study, enabling a unified evaluation of \textbf{robot} and \textbf{environmental} SA using SAGAT across perception, comprehension, and projection, against an information-equivalent mobile baseline.
    \item We found that while AR can improve \textbf{perception-level robot SA} (RQ1a), it does \textbf{not} yield gains in higher-level robot SA, nor does it improve \textbf{environmental SA} (RQ1b) relative to the mobile baseline.
    \item We conducted gaze-based analysis showing that AR's effect on \textbf{saccade velocity} mediates its benefit for \textbf{robot perception}, and that the explicit attention drawn by AR's conformal visuals may explain the lack of gains in environmental SA (RQ2).
\end{itemize}
\section{Related Work}
\label{sec:related_work}
\subsection{AR for Human-Robot Collaboration}
Recent research has explored diverse approaches to improving HRC, including adaptive industrial robots~\cite{Hostettler_CHI_25_Adaptive_Ind_Robots}, behavioral entrainment~\cite{Schneiders_CHI_24_HRI_Entrainment}, and LLM-based speech interfaces~\cite{Padmanabha_UIST_24_Voicepilot_LLMs_HRI}. Complementing these approaches, AR provides a means of embedding robot state, task, and safety information directly within a shared workspace. Prior AR-HRC systems have used spatial visualizations to support safety~\cite{HRI_2023_Robot_SafetyZones_AR, ROMAN_2022_Virtual_Barriers, Choi_VRW_22_XR_Safe_HRC}, trust~\cite{ROMAN_2025_AR_Robot_Trust,SongVR2025VirtualUltrasound}, ergonomics~\cite{Pei_VRW_24_ARHRC_Ergonomics,IRAL_26_Virtual_Model_Control_Handoffs_with_AR}, and understanding of robot state and intent~\cite{ISMAR_23_Convey_Robot_State_Intent,ICRA_2022_Convey_Arm_Intent,fronk2026artoo}. AR has also improved performance in collaborative tasks including deliveries~\cite{ROMAN_2021_AR_Improve_HRC_Efficiency,ICRA_2021_ARROCH_AR_Robot_Collab_Indoors_TabletBased}, sorting~\cite{IROS_2023_Visualization_Effect_On_Eff}, and robot programming~\cite{UIST_2022_AR_Program_TTRobots,UIST_2024_AR_Robot_Training,HRI_26_AR_Feedback_for_Robot_Teaching}.

Of particular relevance to our investigation is AR for collaborative human-robot exploration and search. Chen et al.~\cite{ICRA_2024_MR_Robot_drag_and_drop_quadruped} present an AR interface that aids exploration, while several AR systems developed for SAR tasks~\cite{SSRR_2018_ComeSeeThis,SSRR_2024_AR_For_Improved_SA_UAV,arxiv_2025_STARC_Emerg_Response} aim to enhance SA. However, SA was either considered to be directly ``provided'' by the interface as ``additional information'' or was evaluated through subjective feedback without rigorous measures. Moreover, these studies lack salient events representative of real-world scenarios and do not comprehensively evaluate both robot and environmental SA. Quinn et al.~\cite{quinn2025augmented} likewise elicit SAR experts' preferences for out-of-view hazard visualizations in a field study and derive design guidelines, but stop short of measuring the awareness those visualizations produce. These limitations leave open whether AR improves the robot and environmental awareness needed for safe, effective HRC because of its spatial presentation, or simply because it provides more information. We address this by comparing AR to an information-equivalent mobile baseline in a real-world SAR-inspired task, measuring SA toward both the robot and the environment across perception, comprehension, and projection (RQ1).

\begin{figure*}[t]
	\centering
	\includegraphics[width=0.9\linewidth]{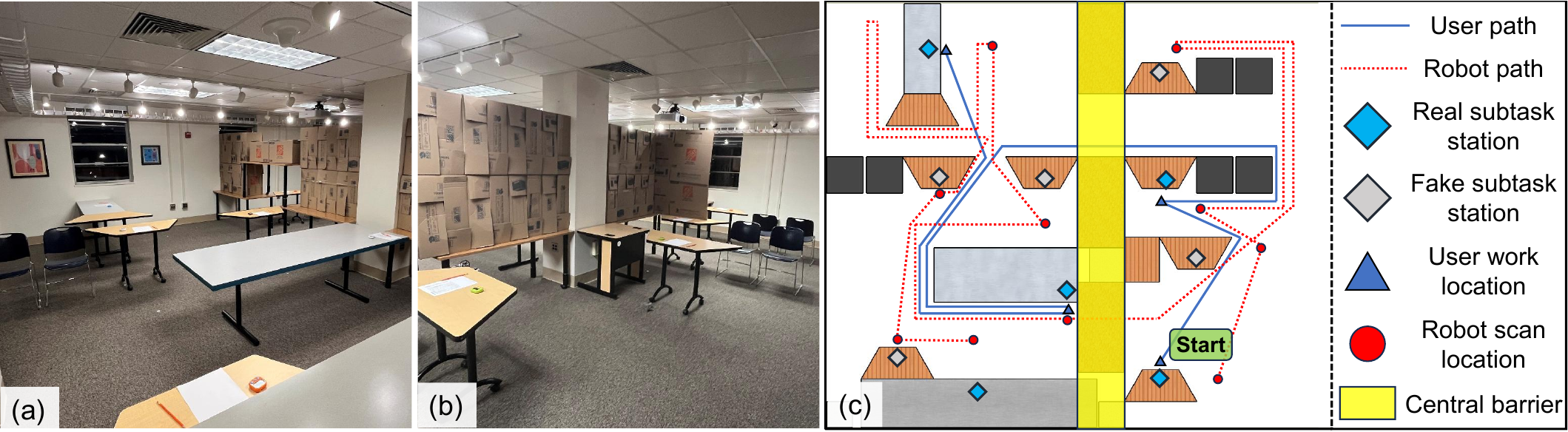}
	\caption{Study space layout and task design. (a, b) Physical layout of the two halves of the room. (c) The 2D view with polygons representing the rectangular and trapezoidal tables. Real and fake subtask stations are marked with blue and gray diamonds, respectively. The user's path to the subtask stations is shown as a solid blue line, while the robot's path to scan and identify subtask stations is shown as a dashed red line. The user's path between station 4 and 5 is omitted for clarity.}
	\Description{A three-part diagram showing the experimental environment. Images (a) and (b) are photographs of the physical study space, showing cluttered office environments with desks, chairs, and boxes serving as obstacles. Panel (c) is a top-down 2D floor plan of the combined space. Blue diamond icons mark ``real'' subtask stations (labeled 1 through 5), while gray diamonds mark ``fake'' stations. A solid blue path illustrates the user's sequential navigation route between the stations, and a dashed red path traces the robot's scanning route throughout the room.}
  \label{fig: room-layout}
\end{figure*}

\subsection{Situational Awareness and Its Evaluation in AR}
Situational awareness (SA) is defined as the perception of elements in the environment, the comprehension of their significance, and the projection of their future states~\cite{endsley1995toward,munir2022situational}. \textit{Endsley's Model of SA}~\cite{endsley1995toward} delineates three hierarchical levels:
\begin{itemize}[itemsep=0pt, topsep=0.3pt, parsep=1pt, leftmargin=0.10in]
\item \textit{Perception} (Level 1): the detection and recognition of the status, attributes, and dynamics of relevant elements in the environment.
\item \textit{Comprehension} (Level 2): the integration of Level 1 elements into a coherent understanding of the environment, enabling the interpretation of the significance of objects and events.
\item \textit{Projection} (Level 3): the ability to anticipate the future states and actions of elements in the environment, based on Levels 1 and 2.
\end{itemize}
While SA is widely acknowledged as critical for decision-making in high-stakes fields like aviation, emergency response, and healthcare~\cite{endsley1999situation,sapateiro2009emergency,kedia2022technologies}, in-depth assessments of SA within AR contexts are still sparse~\cite{woodward2022analytic}. Only a handful of investigations have explored AR-based SA in domains such as industrial tasks~\cite{truong2023study}, aviation~\cite{pan2025situational}, and emergency management~\cite{arxiv_2025_STARC_Emerg_Response}.

This gap is particularly pronounced in AR-HRC. Prior work primarily evaluated robot SA~\cite{Telepresence_2024_AR_Enhance_SA_in_HRI,chacko2026making}, with limited attention to environmental SA. Most of them also relied on coarse-grained measures and did not ground results in established frameworks such as Endsley's model or the Situational Awareness Global Assessment Technique (SAGAT)~\cite{endsley1988situation}, which freezes a task, blanks the display, and queries the operator about perception, comprehension, and projection of the current situation. Other work uses stationary robots or static virtual setups~\cite{ROMAN_2023_Effect_of_AR_on_SA_in_HRC,chacko2026making}, or lacks salient events representative of real-world tasks~\cite{ROMAN_2023_Effect_of_AR_on_SA_in_HRC}. This makes it difficult to determine whether AR improves the broader awareness needed for safe and efficient collaboration or simply enhances robot monitoring while leaving environmental awareness unchanged. We therefore evaluate how AR-HRC affects awareness of both a mobile robot and the environment using structured SA measures and gaze analysis in a dynamic SAR-inspired task. This motivates RQ1a and RQ1b.

\subsection{Gaze Behavior and Attention Allocation in AR}
\label{sec:rw_gaze}
A central promise of AR is that spatially conformal graphics support more effective division of attention between interface content and the physical world than screen-fixed displays~\cite{Wickensetal2022}. Empirical work in driving and assembly contexts suggests that conformal presentation can reduce the cost of switching between near and far domains~\cite{rusch2013directing,bauerfeind2022does}, and AR display placement guidelines aim to minimize attentional disruption during locomotion~\cite{Lee23a,Rzayev18,Manakhov24-GazeOnTheGo}. Eye tracking provides a direct window into these attentional dynamics: fixation and saccade statistics index visual exploration and cognitive load~\cite{liu2022assessing,ledger2013effect,stuart2014quantifying}, while gaze transition and stationary entropy characterize how attention is distributed and moved across areas of interest~\cite{krejtz2014entropy}. However, most evidence for conformal AR's attentional benefits comes from simulators in which virtual content and environment are rendered on the same display, and prior AR-HRC evaluations rarely connect gaze behavior to structured SA outcomes. It thus remains open how optical see-through AR reshapes attention allocation in a real-world HRC task, and whether those changes translate into SA gains (RQ2).

\section{Study Design} 
\label{sec:study_design}
Our user study was designed to answer two research questions:
\begin{itemize}[itemsep=0pt, topsep=0.3pt, parsep=1pt, leftmargin=0.15in]
    \item \textbf{RQ1:} How does a spatially conformal AR interface, compared to an information-equivalent mobile interface, affect users' situational awareness in a dynamic HRC task? Specifically, how does it affect SA toward each of two targets---\textbf{(RQ1a)} the robot and \textbf{(RQ1b)} the surrounding environment---across the perception, comprehension, and projection levels?
    \item \textbf{RQ2:} How does AR reshape users' allocation of visual attention during collaboration, and can these attentional differences explain its effects on SA?
\end{itemize}
To answer RQ1, we measure robot SA and environmental SA using SAGAT freeze probes administered during a dynamic SAR-inspired collaboration task. To answer RQ2, we continuously record participants' eye movements and analyze gaze dynamics, attention allocation, and fixation transitions. 

\subsection{Study Task and Environment}
To emulate robot-assisted SAR scenarios, the overall task was structured with a clear division of labor: the robot's role was to explore the environment and scan for locations requiring human assistance (referred to as ``stations''), while participants navigated to these stations to complete site-specific subtasks (a word search game). Because the robot did not need to wait for a participant to finish a subtask before proceeding, it could continue discovering new stations independently. This parallelization enabled faster task completion than in a single-operator scenario. The division of labor also introduced an intentional attentional demand: participants had to remain aware of the robot's actions and discoveries to coordinate navigation. Furthermore, participants were instructed to imagine operating in a hazardous, partially observable environment (e.g., a collapsed building), where maintaining situational awareness amidst unexpected events is essential for success (see \autoref{sec:study_eval_of_sa}). Although completing subtasks was necessary to simulate individual rescue operations, \textit{the primary focus of this study was the participants' movement between stations}. Participants had to monitor the robot's actions while simultaneously navigating to the next station, effectively necessitating multitasking.

\subsubsection{Study Space Layout}
\label{sec:study_space_layout}
To emulate the conditions of a SAR scenario, we conducted our study in a cluttered, fragmented task space where occlusions and constrained navigation create conditions well-suited for evaluating SA. A $9.22 \times 7.12$~m room was configured as a structured obstacle course (Fig.~\ref{fig: room-layout}(a) and (b)), with tables, chairs, and boxes arranged into a maze-like workspace that limited visibility and movement. Three key features structured the space. First, a central barrier of three tables with stacked boxes aligned with a room column bisected the environment and obstructed inter-side visibility, requiring participants to navigate around it or rely on interface information for situational awareness on the opposite side. Second, a fourth table positioned above a gap in the barrier created an overhead obstacle requiring participants to duck, emulating fallen debris in SAR scenarios. Third, ten tables (rectangular and trapezoidal) were distributed throughout both sides of the environment, serving as potential subtask stations, each holding materials for a word search subtask (e.g., puzzle sheets, timers, and pencils). Overall, this layout created strong visual occlusions, spatially separated task areas, and constrained navigation, requiring participants to integrate information across partially observable regions to maintain SA.

\subsubsection{Task Design}
\label{sec:study_task_design}

The task sequence was as follows: the robot began slightly ahead of the participant by scanning the first station before moving to the next one, while participants navigated to the first station to complete the word search subtask. Each subtask had a predetermined duration between 30 and 60~s, and participants were instructed to focus on the word search during that period. The robot did not wait for task completion and continued exploring to identify subsequent stations. When the timer expired, participants proceeded to the next station indicated by the robot, following the order of discovery. This process was repeated until all robot-designated stations were visited and attempted.

Among the 10 subtask stations, only 5 were ``real'' subtask stations while the other 5 were visually identical but ``fake'', to prevent participants from anticipating which locations they would visit before the robot scanned them. The locations of the subtask stations, robot exploration path, scan locations and user navigation path are shown in Fig.~\ref{fig: room-layout}(c). This task design created a dynamic environment where participants had to continuously integrate information about the robot's discoveries, their own navigation and the environment to successfully complete the joint task, providing a rich context for evaluating both robot and environmental SA.

\subsection{Interface Design}
Given the space layout and task design, both interfaces must serve two roles: (1) providing guidance to the user for navigation to the next subtask station, and (2) supporting monitoring of the robot's state, location and next actions. To enable a controlled comparison of AR and mobile interfaces, we developed both interfaces with equivalent information content and functionality, while leveraging the unique affordances of each platform. 

\subsubsection{Mobile Phone Interface}
\label{sec:mobile_interface_design}

\begin{figure}[h!]
\centering
\includegraphics[width=\columnwidth]
{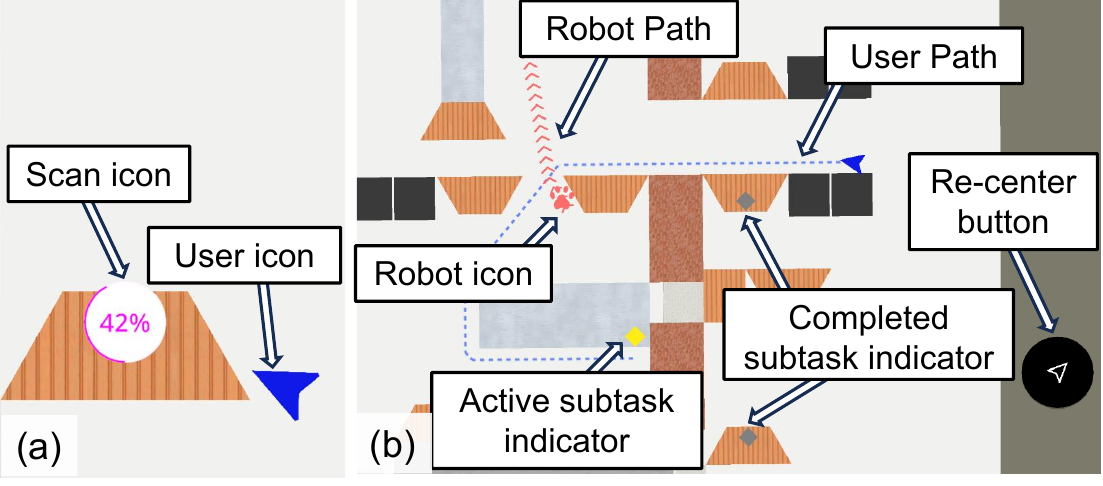}
\caption{Mobile phone interface with a top-down map view showing environment layout: (a) detailed view of the app's interface when the user looks at a subtask station being scanned by the robot; (b) full UI with visuals guiding the user to the next station and facilitating monitoring of the robot's location and planned path. The re-center button resets the map to the default user-centered view.}
\Description{Two mobile phone screenshots showing a 2D top-down map. The interface displays a floor plan with a blue arrow for the user, a blue dashed line for the user's path, a red paw for the robot, and red arrows for its planned path. Image (a) shows the view during a task scan with a magenta progress bar on a trapezoidal table; (b) displays the view when navigating to the next station, with a blue dashed path leading to a yellow diamond marking the next station. A re-center button is visible at the bottom right corner of the screen.}
\label{fig: mobile_interface}
\end{figure}

To provide an effective baseline for AR comparison, we developed a mobile interface inspired by common navigation applications with robot monitoring support. The interface features a 2D top-down map showing the environment layout, obstacles, user position (blue arrow) and user's path (blue dashed line), robot position (paw icon), robot's planned path (red arrows), and subtask stations. An overview is shown in Fig.~\ref{fig: mobile_interface}. When the robot scans a station, a circular magenta progress bar appears at that location; upon completion, a yellow diamond marks valid stations designated as the next target. A dashed blue path shows the navigation route, hard-coded between stations and snapped to the point on the route nearest the user, to reduce jitter from tracking noise. The robot's planned path appears as red arrows extending 4.5~m ahead 
, enabling robot motion anticipation. Completed subtasks appear gray on the map, while pending subtasks appear blue.

Similar to common mobile navigation apps, the app defaults to user-center mode with the map oriented to the user's heading, with a re-center button to restore this view after panning, rotating, or zooming. Users can toggle between landscape and portrait modes. 

\subsubsection{AR Interface}
\label{sec:ar_interface_design}
\begin{figure}[h]
	\centering
	\includegraphics[width=\linewidth]{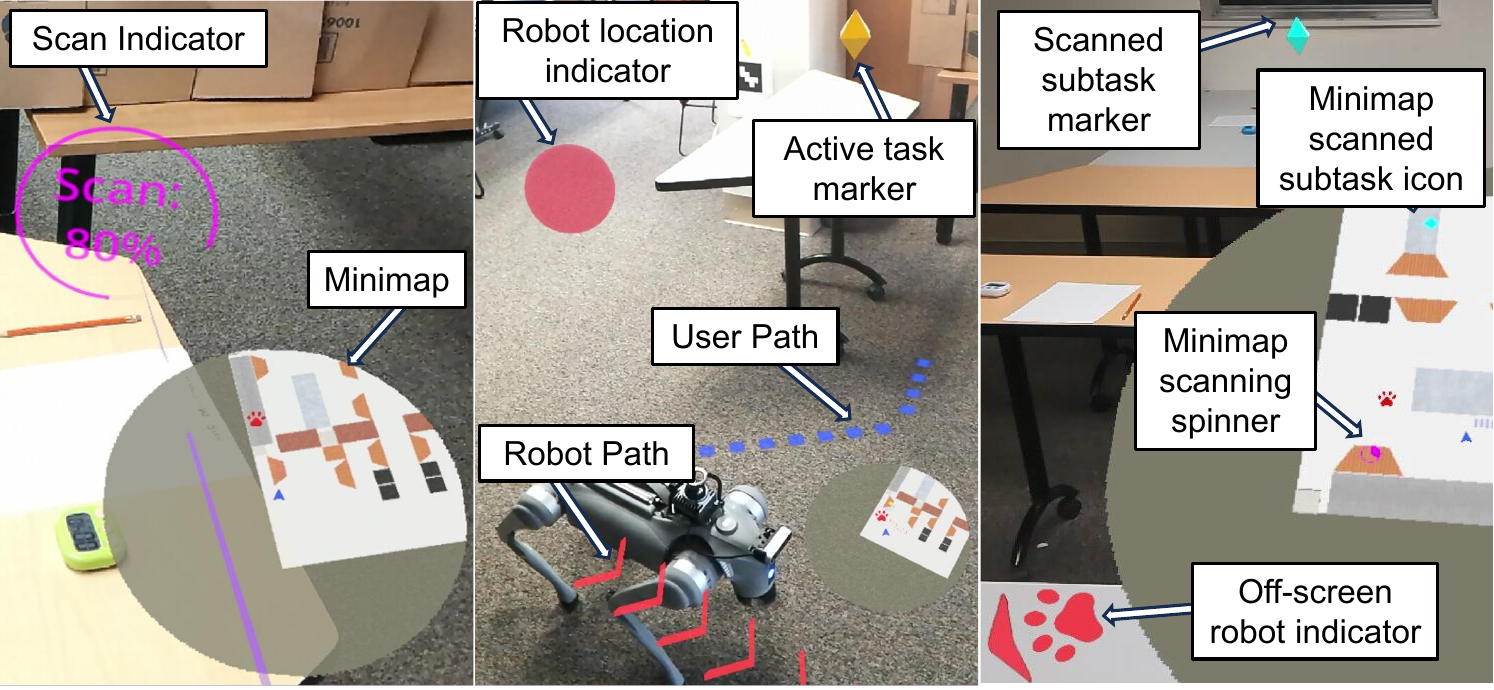}
	\caption{AR interface consisting of conformal visuals, a minimap and an off-screen robot indicator. Note that the robot location indicator (red sphere) remains visible even when the robot is occluded, while the off-screen robot indicator (red paw with an arrow) is visible only when the robot is outside the user's view.}
	\Description{First-person AR headset view of a physical room with projected virtual elements. Conformal visuals include a red sphere floating above a real robot, red arrows drawn on the floor indicating the robot's intended path, and a red indicator arrow anchored at the edge of the field of view pointing toward the off-screen robot. A top-down minimap is visible at the bottom of the user's vision.}
  \label{fig: ar-interface}
\end{figure}

We designed an interface for optical see-through AR glasses (a Magic Leap 2 (ML2) was used in the study) with two main components: conformal visuals that are spatially registered to real-world objects and locations, and a minimap that provides a persistent, top-down overview of the environment and task-relevant information. An additional off-screen robot indicator was included to provide directional cues when the robot was outside the user's field of view (FOV). An overview of the AR interface is shown in Fig.~\ref{fig: ar-interface}.

\noindent\textbf{Conformal Visuals.} 
We developed robot-related conformal AR visuals inspired by prior AR-HRC work~\cite{ISMAR_23_Convey_Robot_State_Intent,ICRA_2022_Convey_Arm_Intent}. A red sphere hovers above the robot, remaining visible even when occluded by physical objects, and scaling with distance to aid depth perception. The robot's intended trajectory is visualized as repeating red arrows on the floor, extending 4.5~m ahead. A red paw icon with a directional arrow anchors to the edge of the ML2's display and rotates to point toward the robot when it is outside the FOV. Together, these indicators enable users to quickly infer the robot's location and intent, reducing the need for continuous direct observation.

We also include conformal visuals to support subtask discovery and navigation. When the robot scans a potential subtask station, a circular magenta scan indicator appears anchored above it. This indicator conveys scan progress through a filling arc and a percentage value (e.g., ``Scan: 80\%''), allowing users to monitor the robot's sensing process at a glance. Upon completion, a yellow diamond marks valid stations as the next objective. To guide users when they navigate between stations, a dashed blue path is rendered along the floor from their current position to the next station. 

To reduce the distracting potential of the AR interface and avoid negatively impacting environmental SA, we visualized the user path not as a dynamic line that continuously updates with user movement, but rather as static waypoints lying on the ground from one station to the next~\cite{Lee22, Renner_Pfeiffer20}, with the traveled portion hidden. In this way, the dashed line stays stable against the ground, reducing the flickering and visual noise that may come with a fully dynamic path. The robot path on the floor is visualized in the same way. 

\noindent\textbf{Minimap.} To support user awareness of the robot and surrounding environment, the AR interface includes a circular minimap anchored to the bottom-right of the user's view (\autoref{fig: ar-interface}). Prior work has shown that minimap and world-in-miniature designs can improve user understanding in AR-guided navigation~\cite{MobileHCI_2011_AR_Indoor_Nav,CHI_23_AR_Nav_Aid_Search_Impacts}, with similar principles applied in HRC exploration tasks~\cite{ICRA_2024_MR_Robot_drag_and_drop_quadruped}. Our minimap aggregates environmental information into a single, stable view, enabling users to maintain global SA even when the robot, task locations, or planned paths fall outside their immediate FOV.


The presentation of the map is identical to the mobile phone interface. The map orientation is always aligned with the user's heading. The minimap UI is placed in the bottom-right corner of the headset's FOV to minimize time spent switching focus between it and the environment during walking~\cite{Lee23a, Rzayev18}, and its movement relative to the head is slightly smoothed for improved legibility~\cite{Manakhov24-GazeOnTheGo}. 

\subsubsection{Implementation}
\label{sec:localization_proc}
~\\
\noindent\textbf{System Architecture.} The AR application was developed using Unity 6000.2.6f2 and deployed on the ML2 headset. The mobile application was developed using Android Studio for a Google Pixel 7. The robot, a Unitree Go2 robot dog, was controlled via ROS2, with a central server facilitating communication between the robot and the AR system. 

\noindent\textbf{Coordinate Frame Alignment.} To ensure accurate spatial registration of the robot across both interfaces, we align the robot's map frame to the AR world frame through two stages. First, we compute a coarse transformation using a shared AprilTag marker observed by both the ML2 headset and robot, with poses transmitted to the server. Second, we refine this transformation by applying point-to-plane iterative closest point (ICP)~\cite{ptp_icp} to align the ML2 spatial map with robot LiDAR scans. This method minimizes distances between corresponding points by projecting them onto target surface tangent planes, yielding accurate rigid transformation estimates. By incrementally incorporating new point cloud observations, we maintain precise registration between AR content and the robot throughout the study. Note that given the necessity of continuously running ICP to maintain accurate robot alignment, the phone app received robot location information from the AR headset (which served as a relay) rather than directly from the robot.

\begin{table*}[t]
\centering
\small
\renewcommand{\arraystretch}{1.0}
\setlength{\tabcolsep}{4pt}
\caption{SAGAT questions asked at the perception (Perc.), comprehension (Comp.), and projection (Proj.) levels.}
\label{tab: sagat_questions}
\begin{tabularx}{\linewidth}{@{}c c >{\raggedright\arraybackslash}X >{\raggedright\arraybackslash}X >{\raggedright\arraybackslash}X@{}}
\toprule
Target & Level & \textbf{1. Remote-Control Car \& Robot Cross Path} & \textbf{2. Overhead Obstacle \& Robot in the Other Half} & \textbf{3. Human Bypass \& Robot Final Heading} \\ \midrule
 & Perc. & In what direction was the RC car moving? & Was there a head-height obstacle in your forward field of view? & Was another person entering the task area? \\
Env. & Comp. & Was the RC car moving toward your immediate walking path? & Did your projected path lead underneath the obstacle? & What was the person doing? \\
 & Proj. & Based on its heading, will the RC car cross your path ahead of you or behind you? & Would you need to adjust your posture (e.g., duck) to avoid a collision? & If the person continues on their path, what object will they encounter first? \\ \midrule
 & Perc. & Where was the robot relative to you when facing the windows? & Was the robot in the other half of the room? & In what direction was the robot moving? \\
Robot & Comp. & Was the distance between you and the robot increasing or decreasing? & Who was closer to the next task location: you or the robot? & Was the robot moving toward areas with or without active task markers? \\
 & Proj. & If you remain stationary, could the robot collide with you? & Which region of the room will the robot go to next? & Which region of the room will the robot go to next? \\
\bottomrule
\end{tabularx}
\Description{Table detailing the specific questions asked during the three SAGAT probes. Rows are divided into Environment and Robot targets and further categorized by level of awareness: perception, comprehension, and projection. Columns align these questions against the three scenarios: 1. RC car and robot cross path, 2. Overhead obstacle and robot in the other half, and 3. Human bypass and robot final heading.}
\end{table*}

\subsection{Evaluation of Situational Awareness}
\label{sec:study_eval_of_sa}
The key metric for this study is SA measured using freeze probes. At three predefined points during the task, SAGAT probes~\cite{endsley1988situation} concurrently evaluated participants' SA regarding both the robot and the environment. Because awareness of the robot and awareness of the surroundings are inextricably linked in SAR contexts, probing them together was necessary to maintain measurement validity. At each probe, the ML2 display was temporarily whited-out and participants were instructed to stop moving; the robot was paused at the same time to freeze scene dynamics. Participants then answered a questionnaire based only on their understanding of the situation immediately before the interruption.

\begin{figure}[th]
	\centering
	\includegraphics[width=0.85\linewidth]{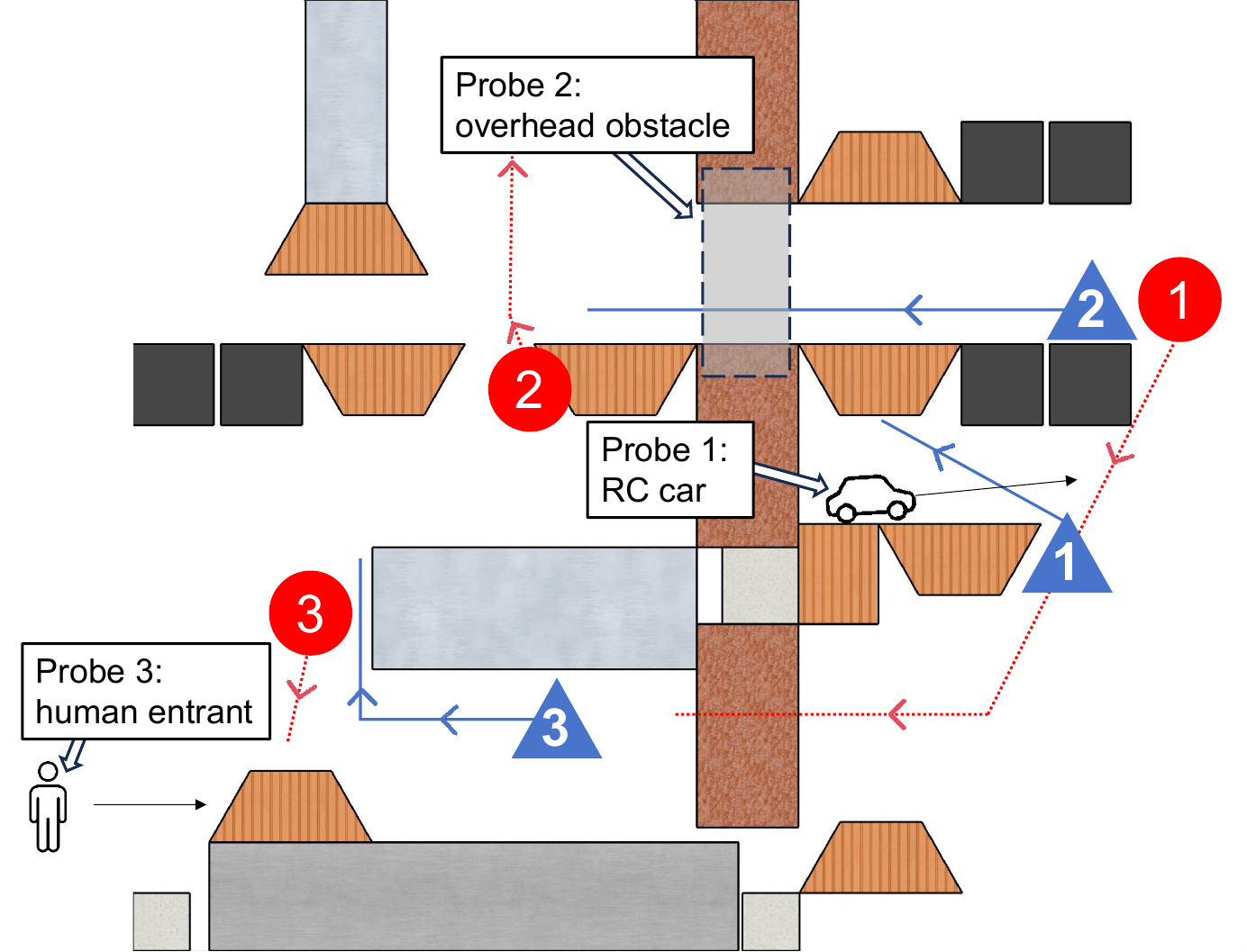}
	\caption{SAGAT probe design. We marked the locations of the user and the robot at the moment of each probe using triangles and circles colored blue and red, respectively. Their heading directions are indicated by arrows on the path. Environmental stimuli relevant to each probe are labeled with icons (RC car, duck, and human).}
	\Description{Top-down floor diagram showing the paths for the user (blue dashed line) and robot (red dashed line) throughout the study. Numbered triangles and circles indicate the exact locations of the user (blue) and robot (red) at three specific interruptions (probes). Icons denote the environmental hazards present at each probe: an RC car crossing at Probe 1, a gray rectangle denoting an overhead obstacle at Probe 2, and a person symbol at Probe 3.}
  \label{fig: sagat-visual}
\end{figure}

Each probe included six questions: three about the environment and three about the robot. Within each set, questions targeted the three SA levels: perception (what is present), comprehension (what it means), and projection (what is likely next). Probes were inserted during transitions between subtask stations when participants were simultaneously navigating and monitoring interface information, with exact locations marked in Fig.~\ref{fig: sagat-visual}. Robot questions covered the robot's relative position, motion, and intended future behavior (e.g., distance trend, heading, next room region). Environment questions covered dynamic hazards and how they intersected participants' paths. Each probe is described below: 
\begin{itemize}[itemsep=1pt, topsep=0.3pt, parsep=1pt, leftmargin=0.15in]
    \item \textbf{Probe 1: Remote-Control Car and Robot Cross Path.} While participants navigated from the first to the second subtask station, a remote-control (RC) car suddenly emerged from behind a table and crossed their path, an unexpected dynamic obstacle (Fig.~\ref{fig:teaser}(a)). At this moment, the robot had departed from a prior scan location and was heading toward the participant. This configuration required participants to simultaneously evaluate the car's trajectory relative to their own path, track the robot's motion and proximity, and assess the risk of colliding with either.
    \item \textbf{Probe 2: Overhead Obstacle and Robot in the Other Half.} As participants moved toward the third subtask station, the probe was triggered the moment they rounded the corner formed by the chairs, with the overhead obstacle directly ahead (Fig.~\ref{fig:teaser}(b)). At this time, the robot had progressed into the unobserved half of the room, where it passed beneath a table. Participants were therefore required to anticipate their path relative to the obstacle and the evasive actions needed to avoid it, while maintaining awareness of the robot's position and task progression.
    \item \textbf{Probe 3: Human Bypass and Robot Final Heading.} After the participant completed the third subtask, a researcher entered the task space nearby, looking at a phone while walking toward a trapezoidal table (Fig.~\ref{fig:teaser}(e)). At this moment, the robot was in close proximity, positioned along the perpendicular edge of the table near the participant, and moving toward another potential subtask station in the remaining unexplored area of the room. The probe was triggered as the robot crossed the midpoint of the table. This scenario required participants to monitor multiple dynamic agents simultaneously and predict their respective motions, intentions, and future interactions.
\end{itemize}
These probes jointly operationalize RQ1a and RQ1b. The exact wording of the questions can be found in Table~\ref{tab: sagat_questions}.  

To ensure repeatability across trials, several aspects of the study were tightly controlled. First, the robot was manually operated by a researcher following a predefined trajectory and timing schedule (subtask durations were adjusted so that the robot occupied designated locations when probes were triggered), eliminating variability from navigation delays and ensuring consistent motion relative to participant progress. Second, robot scanning behavior was centrally controlled via an administrator smartphone application, with scanning indicators and task markers triggered at predefined points along the robot's path to ensure consistent timing of task discovery. Third, all SAGAT probes were manually triggered through the same app at predefined spatial and temporal points.

\subsection{Participants}
Thirty participants (19 male, 11 female; mean age 25.6 years) were recruited and randomly assigned to either the AR or the mobile group ($n=15$ per group). To ensure group balance and reduce predictability, we employed permuted block randomization using randomly varying block sizes of 4 and 6. 19 participants reported using robotic systems once or twice, 4 frequently, 3 infrequently, and 4 never. Within the AR group, 9 participants reported using optical see-through AR once or twice and 6 reported never using it. 

\subsection{Study Procedure}
Upon arrival, participants signed an informed consent form and completed a pre-study survey covering demographics and prior experience with robots and AR. Participants were then introduced to the study task through a written summary and researcher explanations specific to their assigned group. After eye tracking calibration, participants completed a practice questionnaire to familiarize themselves with answering questions through a floating webview interface on the AR headset. Note that mobile group participants also wore a headset during the study to enable accurate location and eye tracking, and to deliver questionnaires in the same fully blocked viewing mode as the AR group. Participants then completed the trial, encountering three SAGAT probes as described above, followed by a post-study survey (see \autoref{sec:study_data_collection}). Each session lasted approximately one hour.

\subsection{Data Collection and Metrics}
\label{sec:study_data_collection}
\noindent\textbf{Situational Awareness.} Situational awareness was evaluated by comparing participants' SAGAT responses to the correct answers for each probe, resulting in a binary score for each question.

\noindent\textbf{Navigation Time.} We collected total navigation time for each participant, which was defined as trial completion time excluding time spent on word searches.

\noindent\textbf{Gaze.} To answer RQ2, using the Magic Leap OpenXR Eye Tracker Feature, we recorded cyclopean gaze direction in world coordinates at 60~Hz. The API identified gaze events: fixations, saccades, pursuits, and blinks. However, given the black-box nature of the algorithm and because pursuits cannot occur without a moving target to track, we re-labeled pursuits using velocity ($100^\circ$/s) and acceleration ($2000^\circ$/s$^2$) thresholds~\cite{stuart2014quantifying}: samples above both were saccades, below both were fixations, and others were undefined. Gaze targets were labeled via Unity raycasting for AR visuals and via post-study egocentric video analysis for the physical robot (both groups) and phone screen (mobile group).

We extracted six gaze segments per participant: three between subtask completion and probes, and three between probes and the next subtask start. Segment boundaries were marked by questionnaire times (probes), gaze leaving subtask objects (subtask completion), or gaze entering the next station (subtask start). Metrics included mean fixation duration (Fix-Dur), fixation rate (FR), fixation ratio (Fix-Ratio), mean saccade amplitude (Sacc-Amp), saccade velocity (Sacc-Vel), and blink rate (BR). To analyze attention, we defined group-specific AOIs: for the mobile group, phone screen and environment; for the AR group, minimap, conformal visuals, and environment. The off-screen robot indicator was included in the conformal visuals as it is spatially registered to the physical robot and serves as a virtual analog of far-domain elements~\cite{rusch2013directing}. We treated the minimap as the AR analog of the phone screen, and grouped conformal visuals with the environment due to their low switching cost~\cite{rusch2013directing,bauerfeind2022does}. Gaze metrics for each AOI included Fix-Dur, FR, Fix-Ratio, fixation share (Fix-Share), and dwell share (Dwell-Share), the proportions of total fixation and dwell time falling on that AOI. We also calculated transition frequency (Fix-Trans-Freq; transitions between consecutive fixations per second), stationary gaze entropy (SGE), and gaze transition entropy (GTE) to quantify visual attention uncertainty (higher SGE indicates more equal distribution; higher GTE, more exploratory behavior~\cite{krejtz2014entropy}). Eye tracking data were unavailable for the first two participants due to hardware issues, so analysis included 28 participants (14 per group). Segment timing was also used to calculate navigation efficiency.

\noindent\textbf{Post-Study Surveys.} After each trial, participants completed the NASA-TLX~\cite{NASATLX} (workload), UMUX-Lite~\cite{UMUX-Lite} (usability, 0--100), and UEQ-S~\cite{UEQ-S} (experience, $-3$ to $3$). Two bipolar slider questions (0--100) assessed information amount in the interface (``far too little'' to ``far too much'') and attentional focus (``100\% on interface'' to ``100\% on physical world''). Participants rated agreement on four statements (5-point Likert; 1 = strongly disagree, 5 = strongly agree): ease of environment tracking around the robot, interface distraction from the environment, agreement that they rarely looked away from the interface, and confidence in noticing unexpected changes. Finally, participants provided open-ended feedback on their experience.
\section{Study Results}
We organize our results around the two research questions. We first verify that the two interfaces provided a fair basis for comparison, then report SA outcomes toward both targets (RQ1), followed by attention allocation and gaze behavior (RQ2).

\subsection{Interface Comparability}
While AR was well received by participants who found it ``cool'', ``easy to use'' and ``helpful'' (as evidenced by subjective feedback), two-sample t-tests on the post-study survey data revealed no significant differences between the AR and mobile groups across several key metrics. Both groups reported similar ratings of the amount of information in the interface ($M_{AR}=45.8, M_{Mobile}=45.07, p=.88$, where 50 means ``just right''), and no significant differences were found in NASA-TLX workload scores (mental: $M_{AR}=58.0, M_{Mobile}=60.3$; physical: $M_{AR}=18.3, M_{Mobile}=23.0$; temporal: $M_{AR}=47.7, M_{Mobile}=54.7$; performance: $M_{AR}=43.4, M_{Mobile}=40.0$; effort: $M_{AR}=52.3, M_{Mobile}=56.3$; frustration: $M_{AR}=39.0, M_{Mobile}=30.7$. All $p \geq .31$). Similarly, UMUX-Lite usability scores were not significantly different ($M_{AR}=84.2, M_{Mobile}=79.2, p=.38$), nor was the UEQ-S score ($M_{AR}=1.33, M_{Mobile}=0.98, p=.11$, score $>$ 0.8 indicates positive evaluation). Navigation efficiency was also not significantly different, with the average time taken to navigate from station 1 to station 4 being $M_{AR}=73.47$ seconds and $M_{Mobile}=82.42$ seconds ($p=.40$). We found no evidence that the interfaces differed in usability, workload, or efficiency, which is consistent with the intended information equivalence of the two interfaces.

\subsection{RQ1: Effect of Interface on Situational Awareness}
Following the practice of prior work~\cite{truong2023study}, we analyzed SAGAT responses using a generalized linear mixed model (GLMM) with a binomial distribution, including fixed effects for interface type (mobile vs.~AR), SAGAT target (environment vs.~robot), and SAGAT level (perception, comprehension, projection), as well as their interactions. We included random intercepts for participants and SAGAT probe number to account for individual differences and probe-specific effects. The model can be expressed as follows:
\begin{center}
    \ttfamily
    Score $\sim$ Interface * SAGAT Target * SAGAT Level + (1 | Participant) + (1 | Probe Number)
\end{center}

Using the \texttt{pymer4} package~\cite{jolly2018pymer4}, we fitted the model and found that random intercepts for participants accounted for a notable portion of the variance (Var = 0.214, SD = 0.463). A boundary (singular) fit warning was observed, as the variance accounted for by the Probe Number random effect collapsed exactly to zero (Var = 0.000, SD = 0.000). This singularity is a common and expected artifact of the small number of levels in the \texttt{Probe Number} grouping factor ($N=3$); however, we retained the random effect structure to reflect the underlying experimental design~\cite{barr2013random}, as retaining a random effect unsupported by the data reduces power but does not inflate Type~I error for the fixed effects~\cite{matuschek2017balancing}.

\begin{figure}[t]
	\centering
	\includegraphics[width=1.0\linewidth]{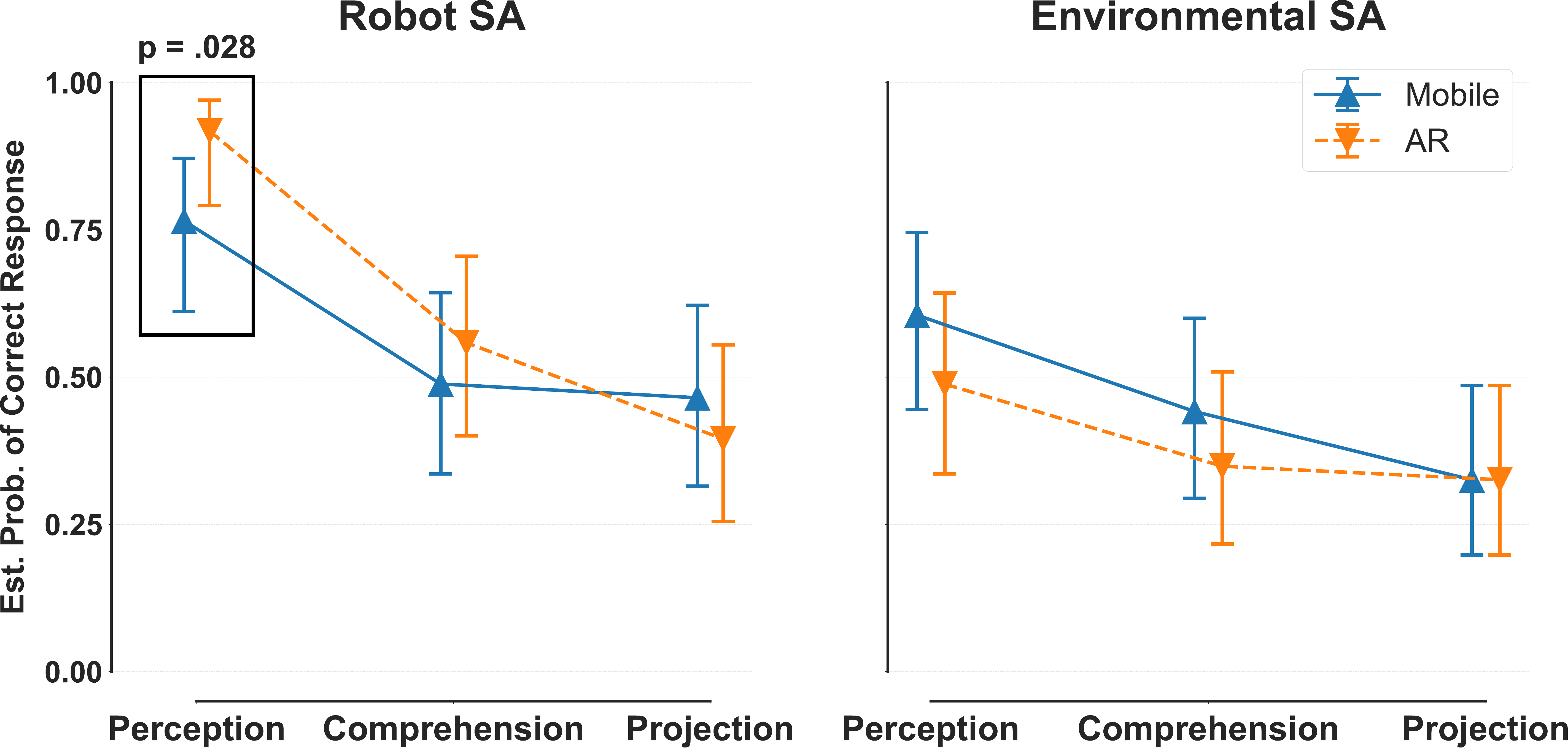}
	\caption{Estimated probability of a correct SA response across three levels, for robot (left) and environmental awareness (right). Points are estimated marginal means (EMMs) from the GLMM; error bars are 95\% confidence intervals adjusted for random effects. The box marks the significant AR advantage in robot perception ($p=.028$).}
  \Description{Two side-by-side line graphs plotting the estimated probability of a correct SA response across three SA levels (Perception, Comprehension, Projection) on the X-axis, ranging from 0.0 to 1.0 on the Y-axis. The left graph (Robot SA) shows the AR condition line starting significantly higher than the mobile condition line at the Perception level (AR at approx. 0.91, Mobile at approx. 0.76), with the gap narrowing at the Comprehension level (AR at approx. 0.56, Mobile at approx. 0.49) and the AR line dropping slightly below the Mobile line at the Projection level (AR at approx. 0.40, Mobile at approx. 0.47), though confidence intervals overlap substantially at this level. Both lines trend downwards overall toward Projection. The right graph (Environmental SA) shows both AR and mobile condition lines closely grouped together between probabilities of 0.3 and 0.7 across all three SA levels, with the mobile condition slightly outperforming AR on perception and comprehension levels before the two lines converge at Projection. Error bars indicate 95\% confidence intervals on all data points.}
  \label{fig: three-way-interaction}
\end{figure}

Analysis of the fixed effects revealed that the main effect of \texttt{SAGAT Level} (specifically \textit{projection} compared to \textit{perception}) had a significant negative impact on the response ($\beta=-1.15, p=.028$), indicating a lower likelihood of a positive outcome at this level. While the main effects of \texttt{Interface} and \texttt{SAGAT Target} were not significant on their own, there was a significant Interface $\times$ Target interaction, with the AR $\times$ Robot contrast at $\beta=1.70, p=.028$. This suggests that AR substantially increased the likelihood of a correct response for questions related to the robot, but not for questions related to the environment.

Moreover, a significant three-way interaction among interface type, SAGAT target, and SAGAT level ($\beta=-1.99, p=.046$) further complicated this effect: the positive effect of AR for robot-related questions was substantially reduced, with point estimates suggesting it may even reverse at the projection level.

\subsubsection{Robot SA (RQ1a)}
We further split the model to analyze the effects on robot and environmental SA separately (for which we removed the \texttt{SAGAT Target} term in the model). For robot SA, analysis of the fixed effects indicated that response correctness was significantly lower at the comprehension ($\beta=-1.23, p=.017$) and projection ($\beta=-1.34, p=.009$) levels. The main effect of the AR interface type showed a marginally significant positive trend ($\beta=1.43, p=.065$), suggesting a potential benefit of AR at the perception level. The interaction term between AR and projection level was also significant ($\beta=-1.78, p=.036$). This indicates that the potential benefit of AR for robot SA is significantly reduced at the projection level.

To further explore the effect of AR on robot SA at perception level, we conducted a post-hoc analysis on 3 perception-level questions related to the robot using chi-square tests (Fig.~\ref{fig: three-way-interaction}). Only in the first probe, which asked about the robot's position relative to the participant, did we find a significant positive effect of AR ($\chi^2(1, N=30) = 5.4, p=.02$), with 13 out of 15 participants in the AR group answering correctly compared to only 7 out of 15 in the mobile group. With respect to RQ1a, AR thus improved robot SA selectively: the benefit was reliable at the perception level but did not extend to comprehension or projection.

\subsubsection{Environmental SA (RQ1b)}
For environmental SA, response correctness was also significantly lower at the projection level ($\beta=-1.22, p=.012$) compared to perception. However, in contrast to robot SA, the main effect of AR and its interactions with test levels were not statistically significant (all $p > .20$; Fig.~\ref{fig: three-way-interaction}). This suggests that for environmental SA, the AR interface provided no significant advantage or disadvantage across any level when compared to the mobile baseline. We found no significant differences between the AR and mobile groups for any of the 9 environmental SA questions when analyzed separately using chi-square tests. With respect to RQ1b, environmental SA showed no advantage for AR at any level; the following section examines the attentional behavior underlying this null result (RQ2).

\subsection{RQ2: Attention Allocation and Gaze Behavior}
\subsubsection{Self-Reported Attention Distribution}
\label{sec:self_report_attention}
We expected AR to make switching attention between the display and the real world easier, and tested this by analyzing the self-reported attention distribution responses with two-sample t-tests. We found a marginally significant difference on the overall attention focus question, with the mobile group leaning more toward the interface end ($M_{Mobile}=27.53$) compared to the AR group ($M_{AR}=40.73, p=.06$). AR users also agreed less that the interface drew attention away from the real environment ($M_{AR}=3.6$) compared to the mobile group ($M_{Mobile}=4.33, p=.050$), together with a marginally significant increase in confidence in noticing unexpected environmental changes ($M_{AR}=3.0, M_{Mobile}=2.47, p=.09$). These results suggest that participants in the AR group may have felt more comfortable distributing their attention between the interface and the physical world, while those in the mobile group may have felt more compelled to focus on the interface.

\subsubsection{Gaze Analysis}
\label{sec:gaze_analyses}
\textbf{Overall Gaze Patterns.} We first analyzed the overall gaze pattern difference between the AR and mobile groups by fitting a linear mixed model for each gaze metric, with fixed effects for interface type and segment number (\autoref{sec:study_data_collection}) and random intercepts for participants. No model showed a significant main effect of segment number, suggesting that gaze patterns were stable across the six segments. Significant main effects of AR were found for Fix-Dur ($\beta=-.063$~s, $p=.007$), FR ($\beta=.611$, $p<.001$), Sacc-Amp ($\beta=1.695^\circ$, $p<.001$), Sacc-Vel ($\beta=23.00^\circ$/s, $p=.014$) and BR ($\beta=-.261$, $p=.004$), with AR users showing lower Fix-Dur and BR but higher FR, Sacc-Amp and Sacc-Vel than mobile users. 

Regarding AOI-specific gaze metrics, 
AR users showed significant increases in fixation rate (FR: $\beta=.758, p<.001$) and fixation ratio (Fix-Ratio: $\beta=.146, p<.001$) on the environment compared to mobile users.
We next compared attention allocation to the ``map'' by analyzing Fix-Share and Dwell-Share on the minimap (AR group) versus mobile screen (mobile group). We observed significant main effects of AR on both Fix-Share ($\beta=-.303, p<.001$) and Dwell-Share ($\beta=-.241, p<.001$), indicating that AR users allocated significantly less time to the minimap than mobile users allocated to the phone, resulting in a higher proportion of attention dedicated to the combination of conformal visuals and the physical environment itself.
Similar trends were observed for fixation switch frequency, as AR users showed lower within-minimap Fix-Trans-Freq compared to mobile users' within-phone Fix-Trans-Freq ($\beta=-0.372, p=.001$), but higher inter-AOI Fix-Trans-Freq ($\beta=0.184, p=.006$) and within-environment Fix-Trans-Freq ($\beta=0.795, p<.001$).

\noindent\textbf{Pre-Probe 1 Gaze Analysis.} To explore the significant robot perception difference between the AR and mobile groups observed in the first probe, we extracted the gaze metrics immediately prior to the first probe from the first segment and conducted two-sample t-tests to examine the difference between AR and mobile users. Significant differences were found for Fix-Dur ($M_{AR}=0.25, M_{Mobile}=0.31, p=.005$), FR ($M_{AR}=2.48, M_{Mobile}=1.93, p<.001$), Sacc-Amp ($M_{AR}=5.31, M_{Mobile}=3.88, p=.001$), Sacc-Vel ($M_{AR}=189.04, M_{Mobile}=167.96, p=.024$) and BR ($M_{AR}=0.39, M_{Mobile}=0.63, p=.003$), indicating more active visual exploration in the AR condition.

\begin{figure}[t]
	\centering
	\includegraphics[width=\linewidth]{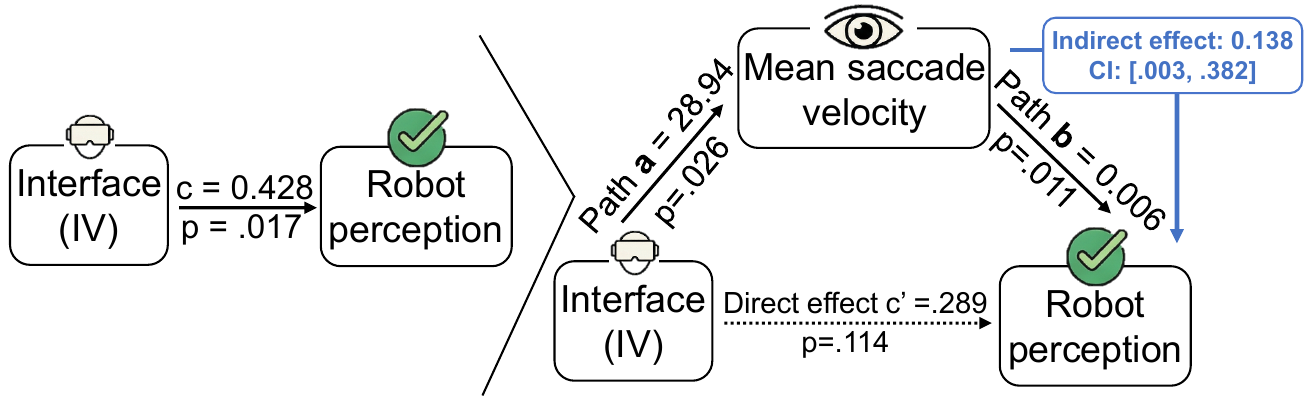}
	\caption{Mediation effect of mean saccade velocity on the relationship between interface type and robot perception in the first probe. The AR interface significantly increased Sacc-Vel, which in turn significantly improved robot perception, fully mediating the interface's effect on performance. ``IV'' stands for independent variable.}
	\Description{Causal mediation model diagram illustrating the relationship between Interface, Mean Saccade Velocity (Sacc-Vel), and Robot Perception. Interface predicts Sacc-Vel ($\beta=28.94, p=.026$), which in turn predicts robot perception ($\beta=0.006, p=.011$). The direct path from Interface to Robot Perception is not significant ($c' = 0.289, p = .114$), indicating Sacc-Vel fully mediates the effect.}
  \label{fig: mediation-illustration}
\end{figure}

\begin{figure*}[t]
	\centering
	\includegraphics[width=0.8\linewidth]{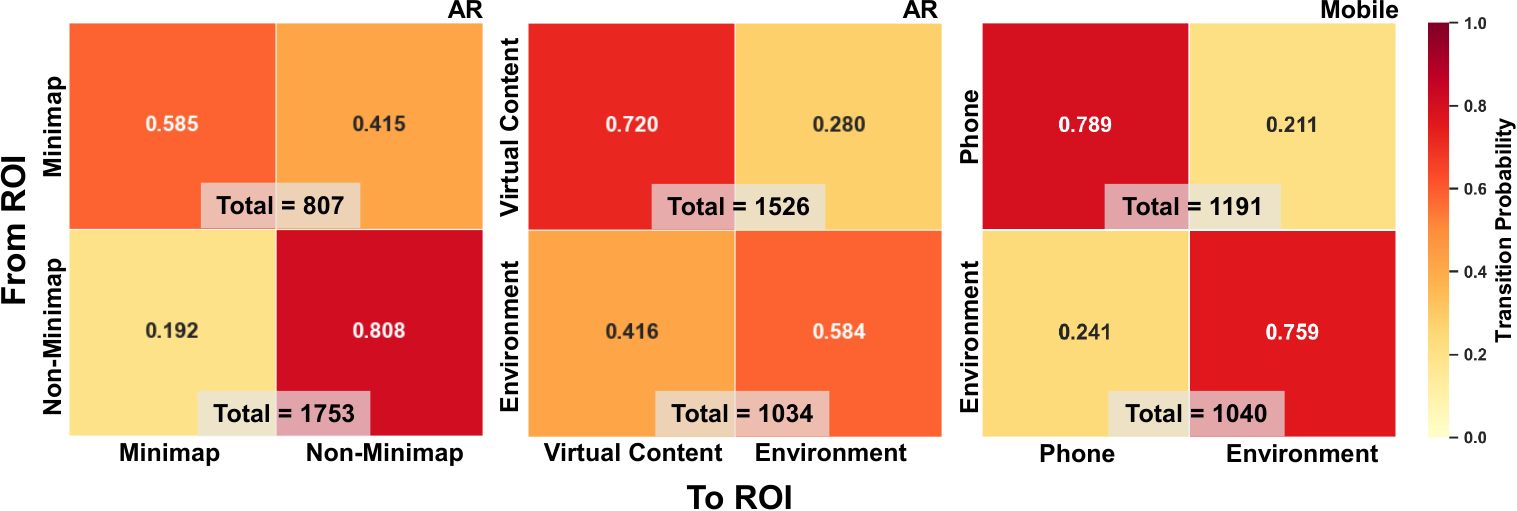}
	\caption{Transition matrix showing the probability of fixation transitions between AOIs for AR and mobile users. The first 2 matrices show the transition probabilities of the AR users, either treating the conformal visuals as part of the environment or as part of the virtual content; the last matrix shows the transition probabilities of the mobile users. The total number of transitions from each AOI (of all users throughout the session) is marked at the bottom of each matrix row.}
	\Description{A heatmap illustrating the transition probabilities between the minimap, phone screen, and physical environment for AR and mobile users. The matrix highlights the frequency of transitions, with redder colors indicating higher probabilities.}
  \label{fig: transition_matrix}
\end{figure*}

\noindent\textbf{Mediation Analysis.} We followed up with a mediation analysis using a bootstrapping approach with 10,000 resamples to examine whether the effect of AR on robot perception in the first probe was mediated by the observed gaze differences. Specifically on Sacc-Vel, the model (see Fig.~\ref{fig: mediation-illustration}) revealed a significant total effect of the interface on robot perception ($c = 0.428, p = .017$). The interface significantly predicted Sacc-Vel ($\beta = 28.94, p = .026$), which in turn significantly predicted robot perception ($\beta = 0.006, p = .011$). Crucially, when accounting for Sacc-Vel, the direct effect of the interface on robot perception became non-significant ($c' = 0.289, p = .114$). The bootstrapped indirect effect was $0.138$, and its 95\% confidence interval did not contain zero ($[0.003, 0.382]$), confirming a significant indirect effect. Taken together, these findings demonstrate full mediation: the improved robot perception observed in AR users is explained by the increased saccade velocity induced by the AR interface.

\noindent\textbf{AOI Analysis.} If the minimap and the conformal visuals are grouped together as the ``virtual content'', AR users had a higher switch frequency within the group ($M_{AR}=1.067$) compared to mobile users' within-phone Fix-Trans-Freq ($M_{Mobile} =0.822$, $p=.017$). Treating the minimap and everything else as two AOIs, AR users' mean SGE was lower than mobile users' ($M_{AR}=0.814, M_{Mobile}=0.959, p=.006$). Merging all virtual content into one AOI eliminated that difference ($M_{AR}=0.925$), while AR users' GTE became significantly higher ($M_{AR}=0.875, M_{Mobile}=0.760, p=.001$). The full transition matrix (see Fig.~\ref{fig: transition_matrix}) further illustrates how this difference occurs. By treating the conformal visuals not as part of the environment but as part of the virtual content (the central part of the figure), we observed not only a higher total number of fixations on the virtual content than on the physical environment for AR users, but also a higher probability of switching from the environment to the virtual content than the corresponding environment-to-phone transition probability for mobile users.
\section{Discussion}
\label{sec:discussion}
Our results give a clear answer to RQ1: AR's SA benefits were selective rather than universal. AR improved perception-level robot SA (RQ1a) but yielded no gains in higher-level robot SA and no improvement in environmental SA (RQ1b). 
Answering RQ2, the gaze data explain this selectivity: AR freed attention from the map but re-invested it in conformal visuals rather than the physical environment, improving robot perception through more active visual exploration while leaving environmental monitoring unchanged. We unpack this account below.

\subsection{AR vs. Mobile: Gaze Pattern Differences and Link to SA}

The phone and AR conditions differed principally in how information was presented. In the phone condition, all task-relevant information --- the next station, the path to it, the robot's location, and its path --- was presented on a single map. In contrast, in the AR condition, these elements were spatially overlaid onto the environment and additionally displayed on a virtual map.

Our results showed that participants had greater perception-level robot SA during the first probe in the AR condition. One of the factors that might have negatively affected awareness during smartphone use was increased cognitive load, which was reflected in higher Fix-Dur and BR values~\cite{liu2022assessing,ledger2013effect}. When using the phone, participants had to interpret map information and then mentally translate it to the environment. In contrast, AR presented navigation information directly within the environment
which may have reduced mental workload, leaving participants with more cognitive resources to track the robot's location when the RC car appeared. We believe the gaze-based indices may be more sensitive to transient load than the retrospective NASA-TLX, which participants completed after a session interleaved with word-search subtasks.

AR users' self-reported comfort distributing attention between the interface and the physical world (\autoref{sec:self_report_attention}) was echoed in their objective gaze behavior. When considering the entire visual field, AR users exhibited more active visual exploration, as evidenced by higher FR, Sacc-Amp, and Sacc-Vel values. In contrast, mobile users displayed more focused gaze patterns, characterized by longer fixations and fewer saccades, likely due to focusing on the mobile screen. As demonstrated by the gaze analysis before Probe 1 (see \autoref{sec:gaze_analyses}), AR users already exhibited more active visual exploration than mobile users, which may have contributed to their better perception of the robot's position.

Mediation analysis showed that the AR benefit in robot perception was fully mediated by increased saccade velocity. We find this result particularly compelling because high saccade velocity implies switching between spatially distant targets in the environment (e.g., from the navigation path to the next station to the robot's path) or between the environment and the map.

\begin{figure}[h]
	\centering
	\includegraphics[width=\linewidth]{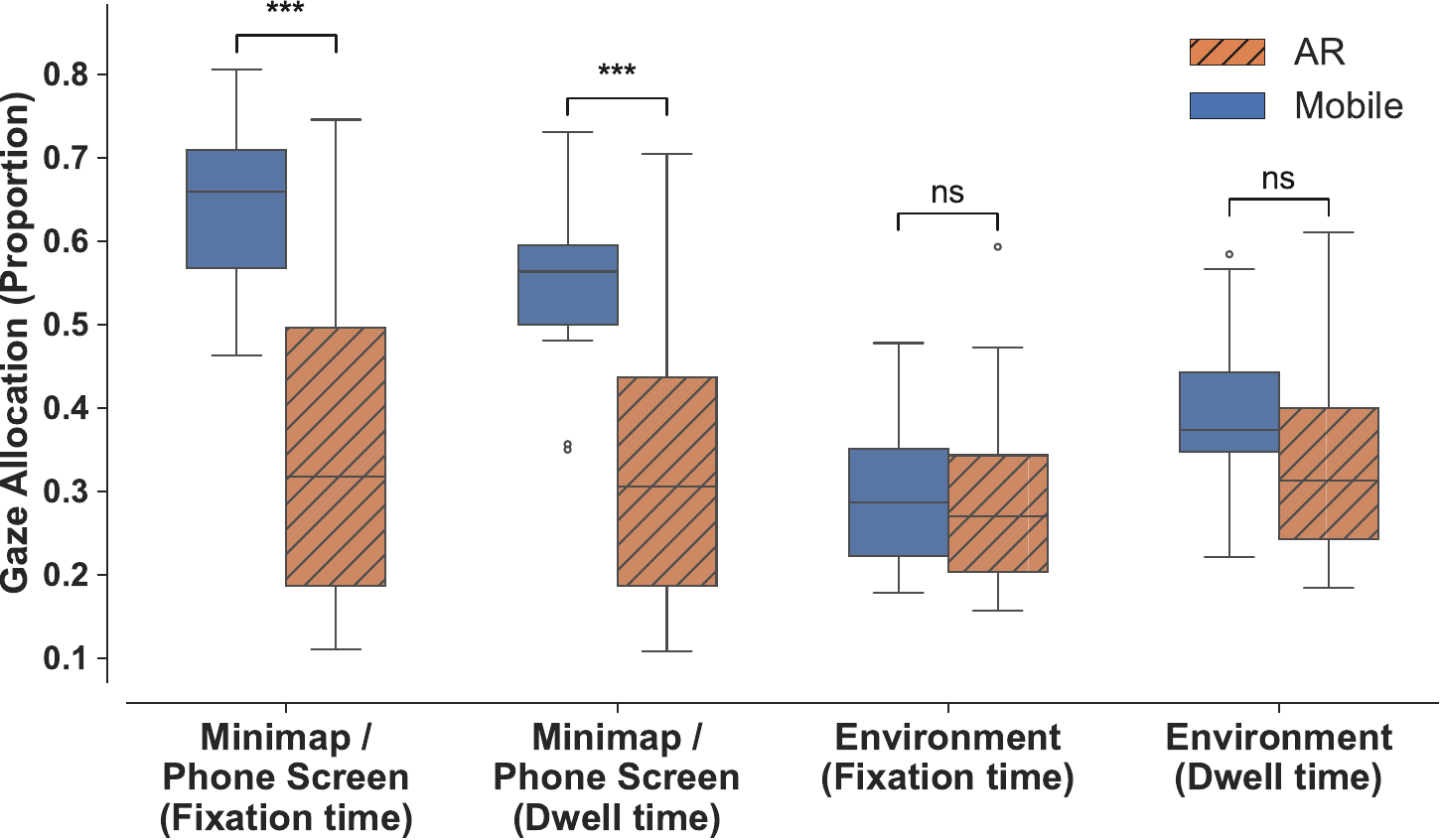}
	\caption{Proportion of fixation time and dwell time allocated to the minimap / phone screen and the physical environment (excluding conformal visuals) for AR and mobile users. AR users allocated significantly less fixation and dwell time to the minimap compared to mobile users' phone screen engagement, but there were no significant differences in fixation and dwell time proportions on the physical environment between groups.}
	\Description{A boxplot chart comparing visual attention allocation between the AR and mobile conditions. Both fixation time and dwell time are plotted for two areas of interest: the minimap/phone screen and the physical environment. The AR group shows significantly lower fixation and dwell time on the minimap compared to the mobile group's phone screen, while both groups show similar proportions of fixation and dwell time on the physical environment, with no significant differences.}
  \label{fig: gaze_explain_env}
\end{figure}

A natural question is why these benefits did not extend to environmental awareness. While two-sample t-tests conducted on the gaze metrics aggregated from the entirety of the session revealed that AR users allocated significantly less fixation time ($M_{AR}=35.0\%, M_{Mobile}=64.7\%, p<.001$) and dwell time ($M_{AR}=32.0\%$, $M_{Mobile}=54.7\%, p<.001$) to the minimap compared to mobile users' phone screen engagement, their fixation time proportion ($M_{AR}=29.8\%, M_{Mobile}=29.4\%, p=.92$) and dwell time proportion ($M_{AR}=33.3\%, M_{Mobile}=39.5\%, p=.17$) on the physical environment (excluding conformal visuals) showed no significant differences (see Fig.~\ref{fig: gaze_explain_env}). In fact, the AR group had on average 28.3\% of fixation time and 26.0\% of dwell time allocated to the AR conformal visuals. The significantly higher fixation rate ($M_{AR}=2.56, M_{Mobile}=1.94, p<.001$) on the environment did not translate to greater information acquisition; rather, it reflected more switching of attention between conformal visuals and the environment. Our AOI analysis supports this (see Sec.~\ref{sec:gaze_analyses}).

To summarize the answer to RQ2: while the level of environmental SA was similar in both the phone and AR conditions, the underlying attentional reasons for this low awareness differed. In the phone condition, participants spent more time looking at the mobile device and less at the environment. This, combined with the increased workload of interpreting map information and mentally translating it to the environment, led to reduced environmental SA. In the AR condition, participants spent more time looking at the environment than in the phone condition; however, their attention was divided across the conformal visuals. They were switching frequently between these visuals and the physical environment, ultimately dedicating a similar amount of time to the physical environment as participants in the phone condition. As a result, environmental SA was comparable across the two conditions.

\subsection{AR Interface Design and Future Work}
When designing the AR interface, we took into account current best practices, such as integrating task-relevant information directly into the surrounding environment~\cite{rusch2013directing,bauerfeind2022does} and providing overview information using a minimap aligned with the surrounding space and presented in the lower right portion of the user's FOV~\cite{Lee23a, Rzayev18}. In both cases, we strove to minimize the attention drawn by the visual elements. For instance, we adjusted the brightness of the conformal visuals and ensured that the navigation paths for both the robot and the user did not create unnecessary movement against the background. We intended for participants to rely primarily on peripheral vision---for example, to follow navigation to the next station---while actively scanning the environment for hazards. Similarly, we carefully tuned the size, opacity, and brightness of the virtual map. While the study results demonstrated that \textbf{the minimap design is viable}---participants used the minimap but attended to it much less than in the phone condition---despite our efforts, the conformal visuals still attracted more attention than intended.

While these results might seem contradictory to findings supporting the advantage of conformal visuals in parallel processing~\cite{rusch2013directing,bauerfeind2022does}, we note that those findings come primarily from simulators, where virtual content and the environment are both rendered. In our real-world study, the AR visuals and the environment were clearly separated, and we expect this separation to persist with current optical see-through hardware.
Therefore, given the limitations of current optical see-through displays, \textbf{conformal visuals should be used more sparingly}. For instance, instead of a continuous navigation path~\cite{Renner_Pfeiffer20}, navigation instructions could appear as discrete billboards at key points along a complex path, freeing attention in between. However, this design requires further investigation.

The broader consideration, however, is the evolving experience of users with AR. As users become more accustomed to conformal navigation guides, we hypothesize that these elements will require less explicit visual attention and can be monitored peripherally. The attention freed in this way would then be available for environmental monitoring, raising environmental SA. This is particularly crucial for real-world applications like search-and-rescue scenarios, where \textbf{rigorous training with the AR interface can foster the necessary expertise for effective environmental awareness and successful HRC}. Future work should investigate this hypothesis through longitudinal studies examining how gaze patterns and SA evolve with increased AR interface experience.

\subsection{Study Limitations}

Our SAGAT probes were designed to assess SA at three levels, and the results showed that participants did perform significantly better at the perception level than at the comprehension and projection levels, which is consistent with the theoretical framework of SA. This suggests that our probes were effective in capturing the intended constructs. The room layout simulating a search-and-rescue scenario also created a non-trivial environment for testing SA, with varying levels of environment and robot SA observable across the three probes. However, we acknowledge potential limitations in generalizability. For instance, the second probe (determining whether the robot was in the other half of the room) was relatively easy, while the third probe occurred with the robot in close proximity, where auditory cues provided strong localization information. This suggests that probe wording may require tuning and that future studies would benefit from a larger physical space to better isolate SA components.

The Magic Leap 2's visor affords only about 108$^\circ$ of unobstructed diagonal view of the real world, well below normal binocular vision. This restricted peripheral vision and may have hindered environmental awareness. However, using the same headset for both conditions controlled this factor. The limited FOV arguably resembles conditions in real SAR, where protective gear restricts vision. Future work should examine whether AR devices with broader FOV (e.g., AR glasses) further enhance SA. Additionally, participants wearing the headset while using a phone may have experienced discomfort; however, we received no spontaneous discomfort reports. Alternative probe delivery methods (beyond full white-out) could improve ecological validity in future studies. 
\section{Conclusion}
\label{sec:conclusion}
In this work, we asked whether a spatially conformal AR interface improves situational awareness (SA) of a robot collaborator and of the environment relative to an information-equivalent mobile interface (RQ1), and how AR reshapes the visual attention underlying SA (RQ2), in a real-world, search-and-rescue-style user study. We found that a custom AR interface significantly improved perception-level robot SA compared to a mobile baseline. Answering RQ2, gaze analysis revealed that this benefit was mediated by more active visual exploration. However, AR's perceptual benefits did not extend to higher-level SA (comprehension and projection) or environmental SA. These findings underscore that conformal visuals in AR do not automatically translate to better environmental monitoring, as freed attention was not redirected to the environment; future work on AR-HRC interfaces should consider minimalist conformal designs that attract less attention if environmental SA is a concern. 

\section{Acknowledgments}
This work was supported in part by NSF grants CSR-2312760, CNS-2112562, and IIS-2231975, NSF CAREER Award IIS-2046072, NSF NAIAD Award 2332744, a Cisco Research Award, a Meta Research Award, Defense Advanced Research Projects Agency Young Faculty Award HR0011-24-1-0001, and the Army Research Laboratory under Cooperative Agreement Number W911NF-23-2-0224. The views and conclusions contained in this document are those of the authors and should not be interpreted as representing the official policies, either expressed or implied, of the Defense Advanced Research Projects Agency, the Army Research Laboratory, or the U.S. Government. This paper has been approved for public release; distribution is unlimited. No official endorsement should be inferred. The U.S. Government is authorized to reproduce and distribute reprints for Government purposes notwithstanding any copyright notation herein. Pavel Manakhov was supported by the European Research Council (ERC) under the European Union’s Horizon 2020 research and innovation programme (Grant No. 101021229 GEMINI).

\bibliographystyle{abbrv-doi}
\bibliography{sample-base}

\end{document}